\documentclass[prx,twocolumn,amssymb,nofootinbib,longbibliography]{revtex4-2}
\usepackage{graphicx, xcolor}
\usepackage{amsfonts, microtype, mathrsfs, bbm}

\usepackage[colorlinks,allcolors=blue]{hyperref}
\usepackage{mathtools, enumerate}
\usepackage{orcidlink}
\usepackage[capitalize]{cleveref}
\usepackage{amsthm}% must be after 

\usepackage{tikz}
\usetikzlibrary{arrows.meta,matrix,calc,positioning}
\DeclareMathOperator{\Tr}{Tr}

\renewcommand{\H}{\mathcal H}

\newcommand{\eps}{\epsilon}

\newcommand{\dagg}{^\dagger}
\newcommand{\norm}[1]{\left\lVert#1\right\rVert}

\renewcommand{\vec}[1]{\mathbf{#1}}

\begin{document}
\title{Estimating applied potentials in cold atom lattice simulators}
\author{Bhavik Kumar\orcidlink{0000-0002-3755-0300}}
\author{Daniel Malz\orcidlink{0000-0002-8832-0927}}
\affiliation{Department of Mathematical Sciences, University of Copenhagen, 2100 Copenhagen, Denmark}

\begin{abstract}
Cold atoms in optical lattices are a versatile and highly controllable platform for quantum simulation, capable of realizing a broad family of Hubbard models, and allowing site-resolved readout via quantum gas microscopes. In principle, arbitrary site-dependent potentials can also be implemented; however, since lattice spacings are typically below the diffraction limit, precisely applying and calibrating these potentials remains challenging. Here, we propose a simple and efficient experimental protocol that can be used to measure any potential with high precision. The key ingredient in our protocol is the ability in some atomic species to turn off interactions using a Feshbach resonance, which makes the evolution easy to compute. Given this, we demonstrate that collecting snapshots from the time evolution of a known, easily prepared initial state is sufficient to accurately estimate the potential. Our protocol is robust to state preparation errors and uncertainty in the hopping rate. This paves the way toward precision quantum simulation with arbitrary potentials.
\end{abstract}

\maketitle

\section{Introduction and motivation} \label{sec:HLfree}

Cold atom quantum simulators aim to implement paradigmatic models of quantum many-body physics, such as the Bose--Hubbard model with precise microscopic control~\cite{Bloch2008,Bloch:2012uep}. A key objective in this context is the ability to realize arbitrary site-dependent optical potentials, that is, to independently tune the on-site energy at each lattice site. Such flexibility enables a broad range of applications in both quantum simulation~\cite{Feynman1982,Georgescu_2014} and quantum information processing---including neutral-atom qubit arrays where addressable local potentials enable site-selective qubit control, programmable problem Hamiltonians, and atomtronic circuits \cite{Weitenberg2011,PhysRevA.75.023615,Amico_2022}. The capability to sculpt spatially complex potentials renders cold atom platforms highly versatile for the study of disordered systems, many-body localization, thermalization, and exotic far-from-equilibrium phenomena such as the quantum simulation of lattice gauge theories~\cite{Schreiber_2015, Choi_2016,Lukin_2019, Mark_2023, Choi_2023,halimeh2023coldatomquantumsimulatorsgauge}. Moreover, applying spatially-varying quasiperiodic potentials has emerged as a powerful tool for probing rich topological phenomena~\cite{Roy_2024,lahiri2024quasiperiodicpotentialinducedcorner}. This is why an increasing number of cold-atom
experiments incorporate spatial light modulators, digital micromirror devices, or other custom optics to
achieve controllable potential landscapes.

While a lot of progress has been made towards single-site controllability, it is fundamentally limited by optical diffraction, which means that potentials can only be realized with finite spatial resolution even in state-of-the-art setups~\cite{Henderson_2009,Weitenberg2011, Zupancic2016, Asteria_2021, Bakr:2010wwy}.
As a result, the following situation arises: the experimenter wishes to apply a certain on-site potential $V$, but instead they apply the potential $W$. The difference $V-W$ is constant over many repetitions of the experiment, i.e., it constitutes a bias (and not fluctuating noise).
The difference between the expected and actual implemented Hamiltonian will lead to a simulation error that accumulates with time, thus preventing high-precision quantum simulation.

\begin{figure}[t]
    \centering
    \begin{tikzpicture}[>=Latex,scale=1.0]
        %----------------------------
        % Parameters
        %----------------------------
        \def\n{5}          % number of lattice sites (0..n)
        \def\dx{1.5}       % site spacing (≈ λ/2)
        \def\amp{0.9}      % potential amplitude
        \def\yshift{-0.2}  % baseline

        %----------------------------
        % Optical potential  ~ cos^2(kx)
        %----------------------------
        \draw[domain={-0.5*\dx}:{(\n+0.5)*\dx}, smooth, samples=300, thick, blue!60]
            plot(\x, {\yshift - \amp*(cos(180*\x/\dx))^2-0.25});

        % Sites + labels + hopping arrows
        \foreach \i in {0,...,\n} {
            \coordinate (site\i) at ({\i*\dx}, {\yshift - \amp});
            \shade[ball color=blue!55!white] (site\i) circle (0.18);
            \node[below=6pt] at (site\i) {\small $v_{\i}$};
        }
        \pgfmathtruncatemacro{\nminusone}{\n-1}
        \foreach \i in {0,...,\nminusone} {
            \pgfmathtruncatemacro{\nexti}{\i+1}
            \draw[->, thick, black!70, bend left=45] (site\i) to (site\nexti);
        }

        %----------------------------
        % QGM above site 3
        %----------------------------
        \begin{scope}
          \coordinate (micBase) at ($(site3)+(0,1.2)$);  
          \def\micScale{1.2}
          \begin{scope}[shift={(micBase)}, scale=\micScale]
            \fill[red!20, opacity=0.4] (-0.35,0.0) -- (0.35,0.0) -- (0,-1.2) -- cycle;
            \draw[fill=black!80, draw=black, line width=0.3pt]
              (-0.35,0) -- (0.35,0) -- (0.25,0.8) -- (-0.25,0.8) -- cycle;
            \draw[fill=black!70, draw=black, line width=0.3pt]
              (-0.18,0.8) rectangle (0.18,1.5);
            \draw[fill=black!50, rounded corners=1pt, draw=black, line width=0.3pt]
              (-0.35,1.2) rectangle (0.35,1.5);
            \draw[fill=gray!20, draw=black!60, line width=0.2pt]
              (-0.20,1.3) rectangle (0.20,1.4);
          \end{scope}
          \draw[red!60, dashed, line width=0.7pt] (micBase) -- (site3);
        \end{scope}
        \node[black!70] at ($(micBase)+(0.0,2.2)$) {\scriptsize QGM};
        \draw[red!70, ->, line width=0.9pt] ($(micBase)+(0.0,0.1)$) -- (site3);
        \draw[thick, dashed, red] (site3) circle (0.6);

        \coordinate (compBase) at ($(site1)+(0.2,1.1)$);

        \begin{scope}[shift={(compBase)}, scale=0.80] % smaller & closer
          % bezel
          \draw[rounded corners=1pt, fill=gray!10, draw=black] (-1.8,0.6) rectangle (1.8,2.5);
          % screen
          \draw[fill=black!92, draw=black] (-1.65,0.75) rectangle (1.65,2.35);
          % inner frame
          \draw[white, line width=0.25pt] (-1.45,0.95) rectangle (1.45,2.15);

          % --- Scatter plot of C_{ii}(t) on the screen ---
          % Axes on screen (subtle)
          \draw[white!70, line width=0.2pt] (-1.45,0.95) -- (-1.45,2.15);
          \draw[white!70, line width=0.2pt] (-1.45,0.95) -- (1.45,0.95);

          % Scatter points (example samples)
          \foreach \x/\y in {-1.30/1.05, -1.05/1.25, -0.80/1.55, -0.30/1.45,
                             -0.05/1.20,  0.20/1.55,  0.45/1.95,  0.70/1.60,  0.95/1.30,  1.20/1.55} {
            \fill[white] (\x,\y) circle (0.06);
          }

          \node[white, anchor=north west, scale=0.6] at (-1.52,2.12) {$C_{ii}(t)$};
          % stand
          \draw[fill=gray!60, draw=black] (-0.40,0.40) rectangle (0.40,0.60);
          \draw[fill=gray!50, draw=black] (-1.00,0.20) rectangle (1.00,0.40);
        \end{scope}
    \node[anchor=west, inner sep=0pt ]
  at ($(micBase)+(1.0,1.0)$) {\includegraphics[width=0.28\linewidth, keepaspectratio]{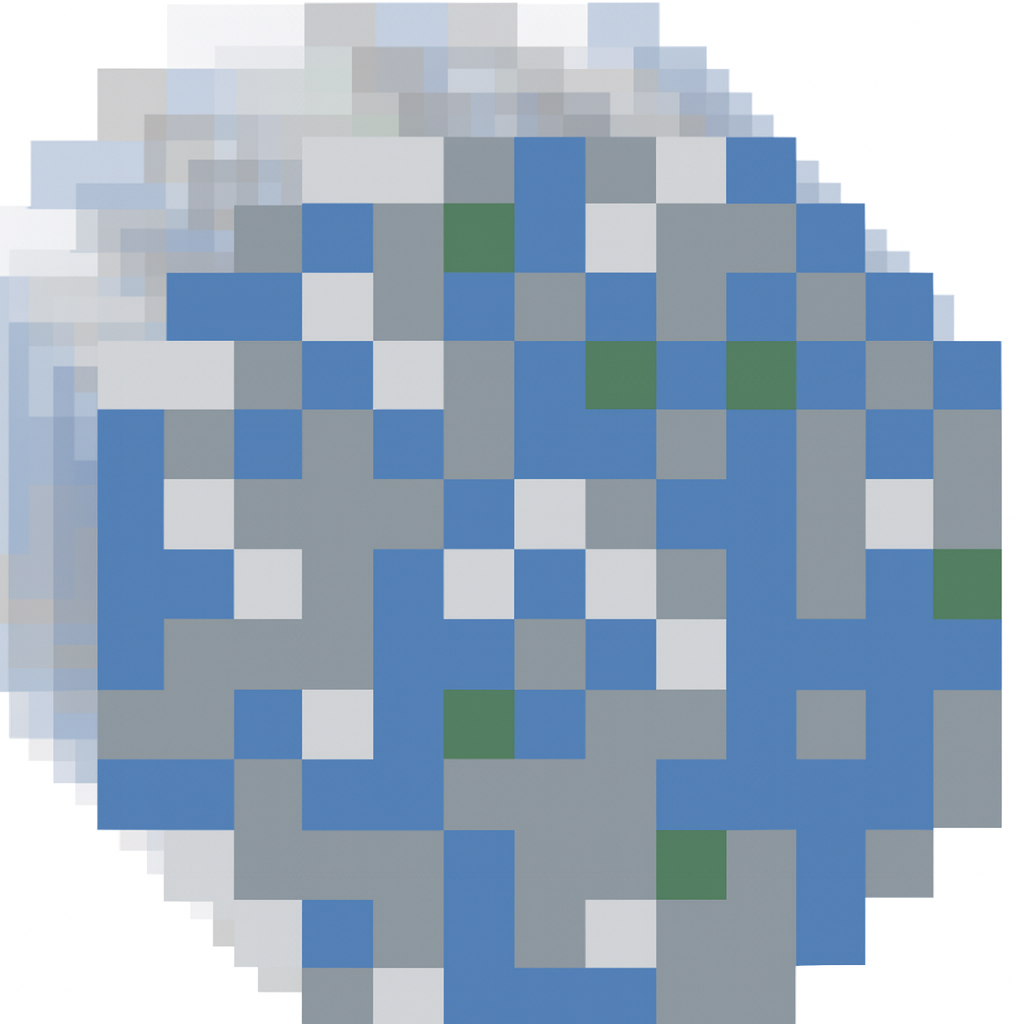}};
 \node[black!70] at ($(micBase)+(2.2,2.7)$) {\scriptsize Imaging};
    
    \label{fig:optical-lattice-minima}
    \end{tikzpicture}

    \caption{Schematic of an optical lattice formed by counter-propagating lasers, generating periodic wells, with on-site energies $v_i$. The atoms can hop from one site to the next. At the end of the experiment, a quantum gas microscope (QGM) is used to read out the location of all the atoms in the lattice, which yields a snapshot.
    By running the evolution for different times, we can infer the applied potential from the snapshots.
    %The dynamics are encoded in the time-evolved correlation matrix $ C_{ij}(t) $. Arrows represent hopping of strength $h$ and the red label marks a measurement site used to collect time-resolved observables $C_{ii}(t)$ acquired by the quantum gas microscope for inference of the potential profile.}
    }
    \label{fig:optical-lattice}
\end{figure}
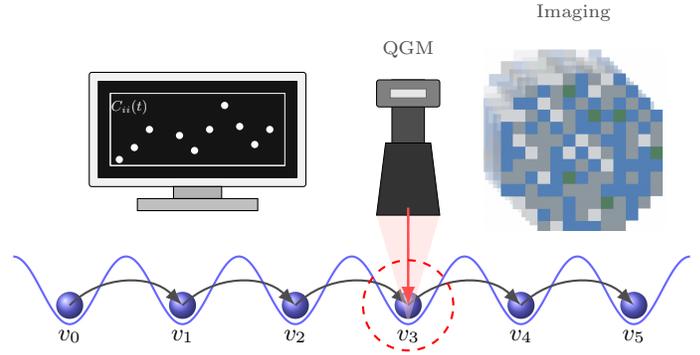

This roadblock can be overcome by learning the Hamiltonian.
A sufficiently efficient and robust algorithm to learn the actually implemented potential could also be used to use iterative feedback calibration~\cite{Koch_2022}, and could be complemented by machine-learning techniques~\cite{Rosi_2013,Cao_2020,reinschmidt2023reinforcementlearningultracoldatom,Wigley_2016}.
There exist a number of proposals to learn Hamiltonians from its dynamics and local measurements~\cite{Bairey_2019, DSF_HL_2024, Valenti_2022, Hincks2018hamiltonianlearningonlinebayesian, Yu_2023, Haah_2024,Li_2020,Burgarth_2009,Villa_2020,Hangleiter_2024}.
These typically employ polynomial interpolation techniques or time series data analysis and require state preparation and measurements in many different local bases.
They are therefore difficult to implement in ultracold atom setups, where the standard measurement is a quantum-gas microscope, which measures the positions of the atoms and is thus in a fixed basis (the number basis).
Recent work has shown that limiting oneself to one measurement basis and fixed Hamiltonians need not necessarily prevent state learning~\cite{PhysRevResearch.5.023027,Tran_2023,PhysRevResearch.6.043118}.
Their key insight is that the scrambling dynamics of the simulator can be used to change the measurement basis. 
One limitation of such an approach is that it often requires one to classically simulate the time evolution of the system, which is difficult for many-body quantum simulators.
Several works have thus developed protocols to extract information from non-interacting quantum dynamics using methods such as convex optimization and (random) free-fermion evolution~\cite{Gluza_2021,Huang_2020,Naldesi2023,Denzler_2024,Wan_2023,PhysRevLett.127.110504,Wu_2024}.

Motivated by this progress, we present a simple protocol to estimate the applied potential (and the hopping) in analogue cold-atom lattice experiments that unifies all favourable properties of previous schemes: (i) implementable without additional experimental modifications (in particular, no local control), (ii) classically efficient postprocessing, (iii) robustness to errors, (iv) constant time evolution ($<10$ hopping times), and (v) low sample complexity.
Taken together, these properties ensure that our protocol is scalable to arbitrary system sizes and that it can be implemented directly in a large number of existing experiments. 
%Here, we ask whether it is also possible to learn the applied potentials solely from quantum-gas measurements and dynamics under the Hamiltonian in question.
To make our protocol classically efficient, we exploit the ability in certain atomic species to turn off interactions via a Feshbach resonance~\cite{Mazurenko_2017,Chin_2010,Bauer_2009}.
Our key result is that time-resolved measurement of diagonals of the correlation matrix $C_{ii}(t)=\langle c_i^\dagger(t) c_i(t)\rangle$ provides sufficient information to reconstruct the actual potential landscape.
Our scheme is well-suited to analog platforms with native access to particle-number readout, requires only easily preparable initial states such as a charge-density wave, and remains robust to realistic preparation errors.

\section{Setting}\label{sec:setting}

For simplicity, we consider a spinless free fermion system on a 1D lattice governed by the quadratic Hamiltonian 
\begin{equation}\label{qHam}
    H = h\sum_{i,j} \left( c_i^\dagger c_j + h.c.\right) + \sum_i v_i c_i^\dagger c_i \eqqcolon \vec c\dagg \cdot \H \cdot \vec c,
\end{equation}
which describes experiments based on cold atoms hopping in an optical lattice.

Here, $v_i$ are unknown onsite potentials and $c_i$ is the annihilation operator acting on site $i$ obeying the canonical anticommutation relations $\{c_{i}^{ },c_{ j}\dagg\}=\delta_{{ i},{ j}}$.
We assume that one can prepare a known initial state with correlation matrix $C(0)$.
The system subsequently evolves unitarily under the Hamiltonian \eqref{qHam}, which yields the correlation matrix
\begin{equation}\label{Prop}
    C_{ij}(t) = \Tr(c_i^\dagger(t) c_j(t) \rho)=e^{i\H^*t}C(0)e^{-i\H^Tt}.
\end{equation}
Readout in cold-atom simulators is performed using a quantum gas microscope, yielding site-resolved snapshots of occupations after imaging, as illustrated in Fig.~\ref{fig:optical-lattice}.
Averaging over $R$ such shots at a number of times $t_j \in [t_0, t_{\max}]$ results in estimates of the local densities
\begin{equation}
	D_{i}(t_j) = C_{ii}(t_j) + \eta_i(t_j), \qquad\eta_i \sim \mathcal{N}(0,\sigma^2),
	\label{eq:data}
\end{equation}
which constitutes our measurement data with random statistical noise of variance $\sigma^2=1/R$.

\section{Learning the applied potential}

We now present two complementary approaches: a rigorous analytic reconstruction, which can be used to certify that reconstruction is possible in principle (that the measurement is informationally complete), and a simple heuristic numerical optimization that achieves much better (realistic) sample complexities.

\subsection{Rigorous protocol}

In the rigorous protocol, our strategy is to fit a polynomial to the measured data $D_i(t_j)$.
From the coefficients of this polynomial one can infer the parameters of the underlying Hamiltonian.
The first time the potential enters the polynomial is at second order,
\begin{equation}
\left.\frac{d^2 C_{ii}(t)}{dt^2}\right|_{t=0} = h^2\beta_i - h\alpha_i \Delta_i + h\alpha_{i-1} \Delta_{i-1},
\label{eq:second-derivative}
\end{equation}
where the potential differences $\Delta_i = v_{i+1} - v_i$, and the coefficients
\begin{align}
	\alpha_i &= C_{i+1,i}(0) + C_{i,i+1}(0), \\
	\beta_i &=4C_{i,i}(0)
	+2\Re\sum_{s=\pm1}(C_{i,i+2s}(0) - 2C_{i-1,i+s}(0)).
\end{align}

\Cref{eq:second-derivative} defines a recursive triangular system, which can be solved for $\Delta_i$ via forward substitution.
For this to succeed, we require that $\alpha_i$ are non-zero. This implies that the currents in the initial state of the protocol must be non-zero.
A potential way to prepare such a state is through a short-time quench from a charge-density-wave state in the presence of a global linear potential, which can be applied with high precision~\cite{Bloch2008,Trotzky2008}.

To estimate the second derivative of the time-evolved correlation matrix entry $ C_{ii}(t) $ at $ t = 0 $, we apply robust polynomial interpolation as introduced in Ref.~\cite{kane2017robustpolynomialregressioninformation} and recently adapted for Hamiltonian learning in Ref.~\cite{DSF_HL_2024}.

	 We pick the times $\{t_j\}_{j=1}^S$ at which we measure at random, distributed according to the Chebyshev measure on $[t_0,t_{\mathrm{max}}]$, which concentrates samples near the interval endpoints and thereby minimizes the worst-case interpolation error. Then, following Ref.~\cite{DSF_HL_2024} we choose $t_0=1/d^2$ and $t_{\mathrm{max}}=t_0+2$, in units of $h$, where $d=O( (\log(1/\eps))^2)$.
	Sampling $S\approx 4d \log(100d)$ time points to some accuracy $\sigma$, robust polynomial regression~\cite{kane2017robustpolynomialregressioninformation} yields a polynomial $\hat p_i(t)$, for which
	\begin{equation}
		\sup_{t\in[t_0,t_{\mathrm{max}}]}|C_{ii}(t)-\hat p_i(t)|\leq 3\sigma,
		\label{eq:polynomial-regression-bound}
	\end{equation}
	and consequently with high probability the error in estimating the second derivative is below $10\sigma d^4$.
	Although this works in principle, we find numerically that we need $d>8$ to obtain sensible results. Following \cref{eq:polynomial-regression-bound}, we require $\sigma<\eps / d^4$ and correspondingly $R\sim d^8/\eps^2>10^7$ snapshots, making this approach experimentally infeasible. Ultimately the reason for this is the sensitivity of polynomial interpolation to statistical noise.

%This result makes explicit how the sample complexity depends on the desired accuracy $ \epsilon $ and the degree of the polynomial. In this case, the potential $V$ is encoded in the second time derivatives of the local correlations $ C_{ii}(t) $, and their estimation is based on reconstructing $ p''_i(0) $ from experimental data.

\subsection{Efficient, heuristic protocol}
To overcome the noise sensitivity and nontrivial state-preparation requirements of polynomial interpolation, we show here that one can instead directly minimize a cost function via gradient-based optimization, thereby avoiding any computation of higher-order derivatives. We define the cost function as
 \begin{equation}
   K(\vec v; C(0))
	= \frac{1}{S}\sum_{j=1}^S
     \left| C_{ii}(t_j; \vec v, C(0)) - D_i(t_j) \right|^2.
 \end{equation}
Here, $C_{ii}(t; \vec v, C(0))$ is the predicted occupation at site $i$ and $D_i$ the measured data, as introduced above.
For the heuristic protocol, we find that a simple charge density wave is sufficient as an initial state, and we adopt this approach in the remainder of the paper.

We numerically address three questions in the following. 
First, what are the optimal choices for the meta-parameters of the algorithm, namely the maximal evolution time $t_{\max}$ and the number of sampled time points $S$? 
Second, how well does the algorithm perform in practice? Specifically, is the sample complexity reasonable, and is the procedure robust to errors? 
Third, how easy is it to find the global minimum of the cost function?
To answer the first two questions, we initialize gradient descent with the true solution and assume that the nearest minimum is also the global minimum as this is numerically less demanding. 
To address the third question, we initialize with guesses randomly distributed around the true solution, where the randomness reflects uncertainty by the experimenter.
We quantify the success of the recovery by computing the average distance of the estimated potential differences, $\tilde{\vec \Delta}$, from the true potential differences $\vec\Delta$,
\begin{equation}\label{MRE}
	\varepsilon_{\text{MRE}} =  \norm{\vec \Delta-\tilde{\vec \Delta}}_1/N.
\end{equation}

\section{Numerical experiments}
We now systematically analyze the influence of (i) the maximal evolution time $t_{\max}$, (ii) the number of equidistant sampling points $S$ within the interval $[0, t_{\max}]$, and (iii) the total number of samples $M = S R$ and system size $N$ on the reconstruction accuracy. We use the Python library \texttt{scipy.optimize.minimize} with the L-BFGS algorithm for the numerical optimization.
In addition to the ideal case, we include the effects of state-preparation errors modeled by a perturbed initial correlation matrix $C^{\mathrm{true}}(0) = C(0) + \gamma\,\Delta C$ and miscalibration of the hopping by $\Delta h$. The entries of $\Delta C$ are drawn from a normal distribution $\mathcal{N}(0,1)$. 
Throughout all simulations, the hopping rate is fixed to $h = 1$, which sets the time scale.

\subsection{Effect of \texorpdfstring{$t_{\mathrm{max}}$}{tmax} and \texorpdfstring{$S$}{S}}

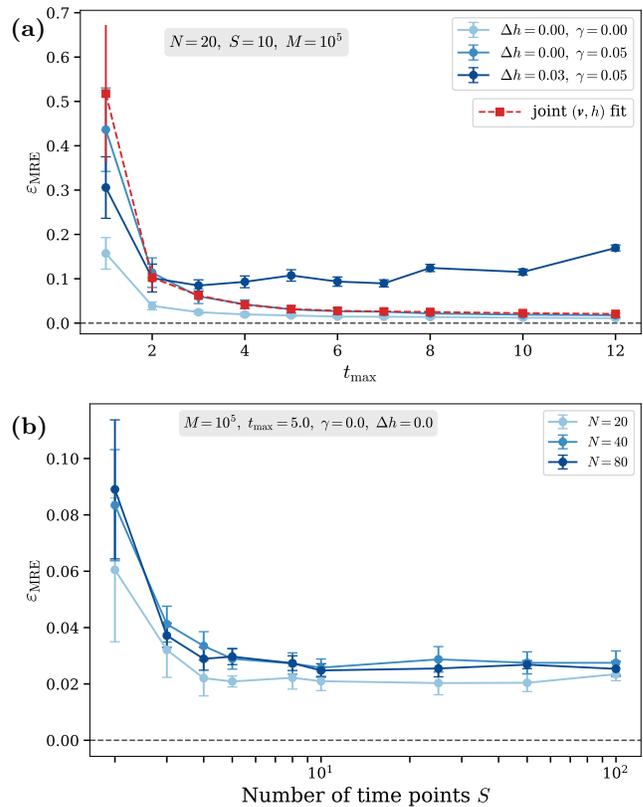
\begin{figure}[t]
    \centering
    \includegraphics[width=0.98\columnwidth]{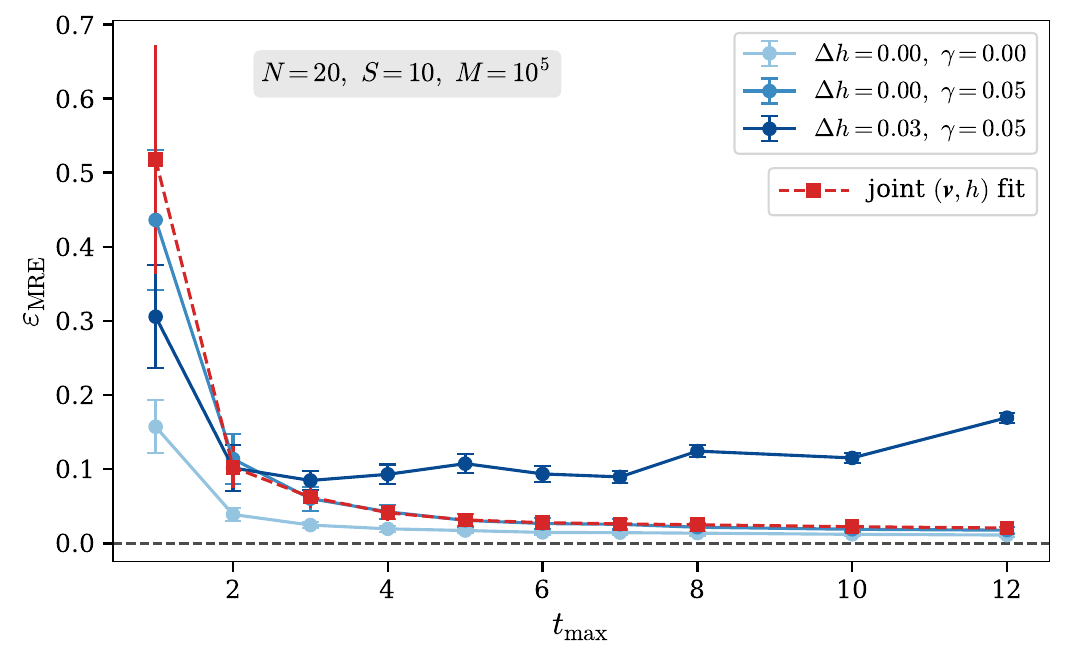}
    \begin{picture}(0,0)
        \put(-245,135){\textbf{(a)}} 
    \end{picture}
    \vspace{0.8em}
    \includegraphics[width=0.98\columnwidth]{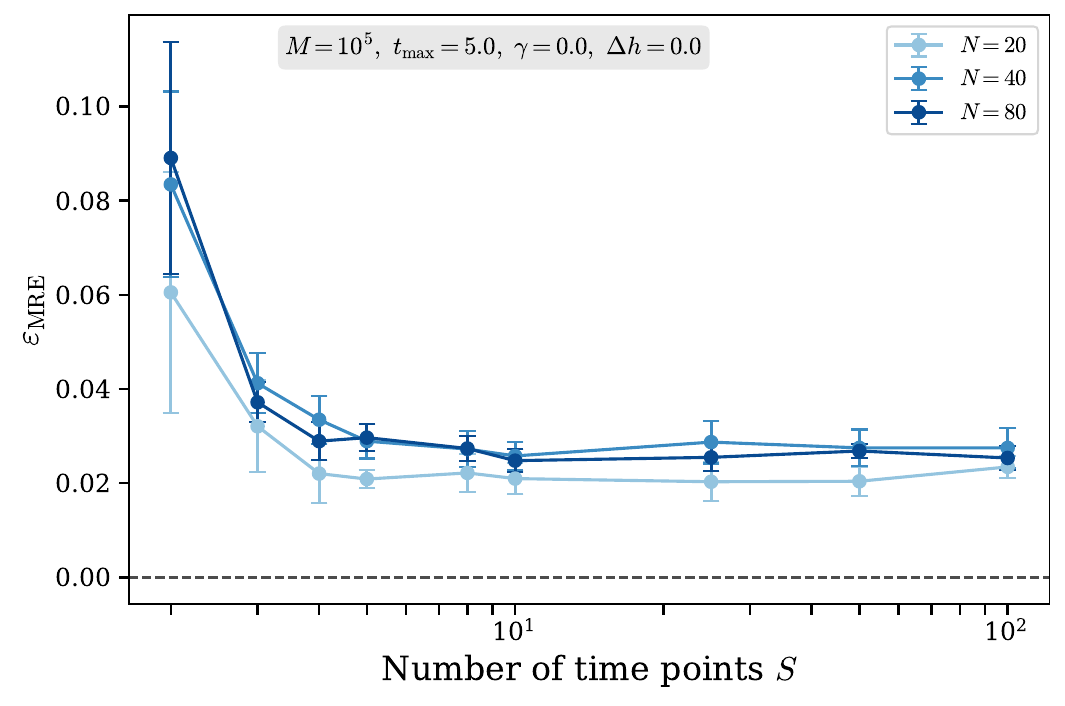}
    \begin{picture}(0,0)
        \put(-245,145){\textbf{(b)}} 
    \end{picture}
    \vspace{-2em}
   \caption{(\textbf{a})~Blue: Mean reconstruction error $\varepsilon_{\mathrm{MRE}}$ [\cref{MRE}] of the estimated potential differences averaged across all sites and many instances of the noise as a function of the maximal evolution time $t_{\max}$ for fixed $S = 10$ and $M = 10^{5}$, and for increasing levels of state-preparation and hopping errors $(\gamma, \Delta h)$. Errors accumulate with time for the dark blue curve due to miscalibration of the hopping strength.
   Red: Additionally fitting the hopping strength removes this error.
   %In the absence of such errors, the reconstruction error decreases monotonically and begins to saturate near $t_{\max} \!\approx\! 5$, yielding diminishing returns beyond this point.At later times, when $\gamma, \Delta h > 0$, state-preparation and hopping errors begin to dominate, leading to an increase in the reconstruction error due to the systematic bias they introduce. This rise in error is primarily driven by the hopping mismatch $\Delta h$; by simultaneously learning the hopping amplitudes, we recover a scaling behavior close to the ideal case, as indicated by the red dashed curve.
   (\textbf{b})~Mean reconstruction error $\varepsilon_{\mathrm{MRE}}$ plotted against the number of sampled time points $S$ for increasing system sizes $N$, with fixed $M = 10^{5}$, and $t_{\max} = 5$. The error decreases rapidly and saturates for $S\geq 5$ at the statistical noise floor determined by $M$ and $t_{\max}$, independent of system size. In both plots, the initial guess for gradient descent is the true potential profile.
}
    \label{fig:MRE_t_S}
\end{figure}

As a first step, we determine the optimal choice of the maximum evolution time $t_\mathrm{max}$ and the number of distinct time points $S$ at which we sample.
As shown in \cref{fig:MRE_t_S}(a), the mean reconstruction error  $\varepsilon_{\mathrm{MRE}}$ [\cref{MRE}] initially decreases rapidly as $t_\mathrm{max}$ is increased, reflecting the increasing sensitivity of the local density dynamics to the onsite potentials. Each data point is averaged over many independent noise realizations.
The mean reconstruction error $\varepsilon_{\mathrm{MRE}}$ begins to saturate near $t_{\max}\!\approx\!5$. 
For finite state-preparation and hopping errors $\gamma,\,\Delta h > 0$, the reconstruction accuracy deteriorates at later times as these imperfections start to affect the dynamics, introducing systematic bias. 
The increase in error is primarily driven by the hopping mismatch $\Delta h$; however, performing a joint optimization over both the on-site potentials $v_i$ and the hopping rate $h$ (red dashed curve) gives near perfect agreement with the error achieved at $\Delta h=0$, showing that this bias can be mitigated. 
We fix $t_{\max}=5$ in the subsequent analyses.

\Cref{fig:MRE_t_S}(b) shows the dependence of the mean reconstruction error on the number of time points $S$ for a fixed number of total snapshots $M=SR=10^{5}$ and constant $t_{\max}=5$, across different system sizes $N$. 
The error decreases rapidly and saturates for $S\!\gtrsim\!5$ at the statistical noise floor determined by $M$ and $t_{\max}$, independent of system size, indicating that sampling at additional time points provides negligible improvement. Going forward, we fix $S=10$.

\subsection{Sample complexity, noise, and system-size scaling}

\begin{figure}[!htb]
    \centering
    % ---------- Panel (a) ----------
    \includegraphics[width=0.98\columnwidth]{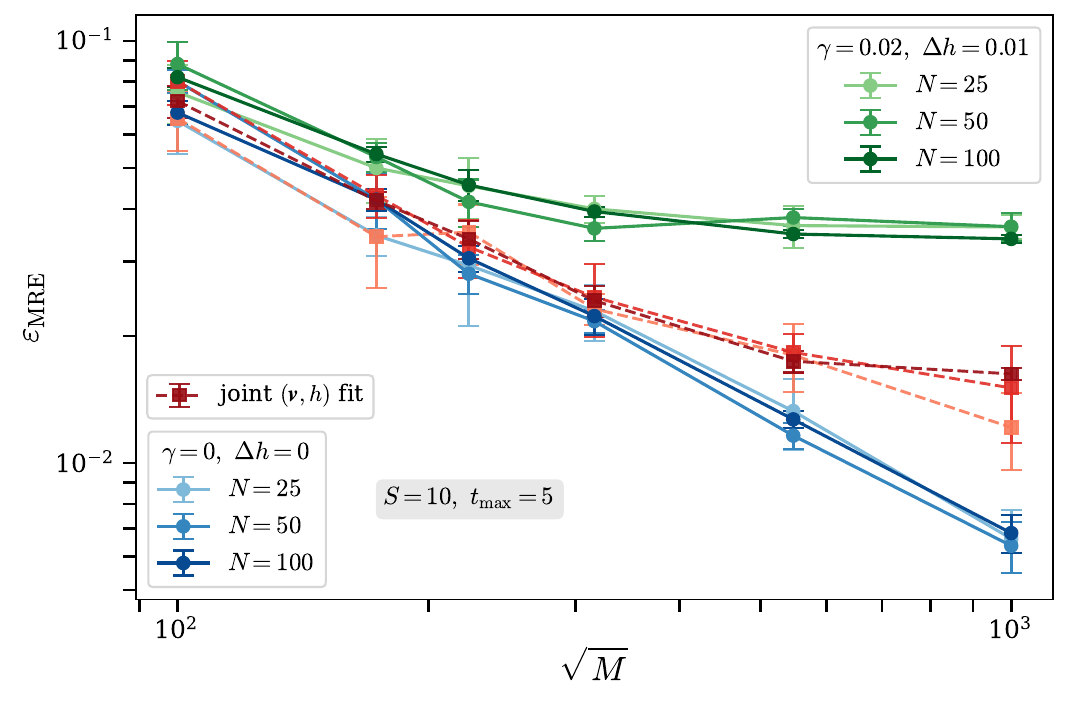}
    \begin{picture}(0,0)
        \put(-245,140){\textbf{(a)}}
    \end{picture}

    \vspace{0.8em}

    % ---------- Panel (b) ----------
    \includegraphics[width=0.98\columnwidth]{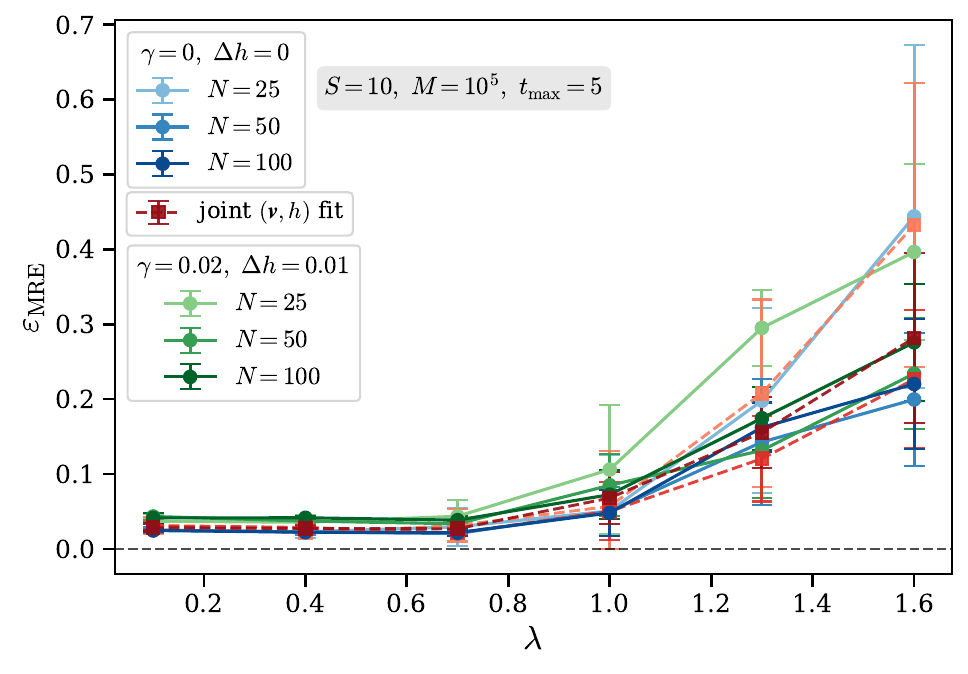}
    \begin{picture}(0,0)
        \put(-245,140){\textbf{(b)}}
    \end{picture}
    \vspace{-2em}

  \caption{(\textbf{a})~Scaling of the mean reconstruction error ${\varepsilon_{\text{MRE}}}$ with the total number of samples $M$ and system size $N$. The error exhibits the expected statistical scaling $\varepsilon_{\text{MRE}}\!\sim\!1/\sqrt{M}$ and remains independent of $N$, confirming the scalability of the learning protocol. In the absence of experimental imperfections (blue curves), the optimizer reliably converges to the true potential profile. When state-preparation and hopping errors $(\gamma,\Delta h)\neq(0,0)$ are introduced (green curves), the optimization still converges but saturates at a biased minimum of the loss landscape. The red dashed curve shows the performance when both on-site potentials and hopping amplitudes are jointly optimized
% , which further suppresses the residual bias and improves overall accuracy. 
    The true potential is generated as $v_i+\lambda\,\Omega_i$, with the optimizer initialized at $v_i$ and $\lambda=0.5$. (\textbf{b})~Dependence of ${\varepsilon_{\text{MRE}}}$ on the initialization uncertainty ${\lambda}$. Convergence to the true potential is maintained for $\lambda\!\lesssim\!1$, demonstrating that the optimization remains robust against substantial initialization mismatch. At larger $\lambda$, the errors grow as the initial guess departs too far from the true potentials.}

    \label{fig:MRE_M}
\end{figure}
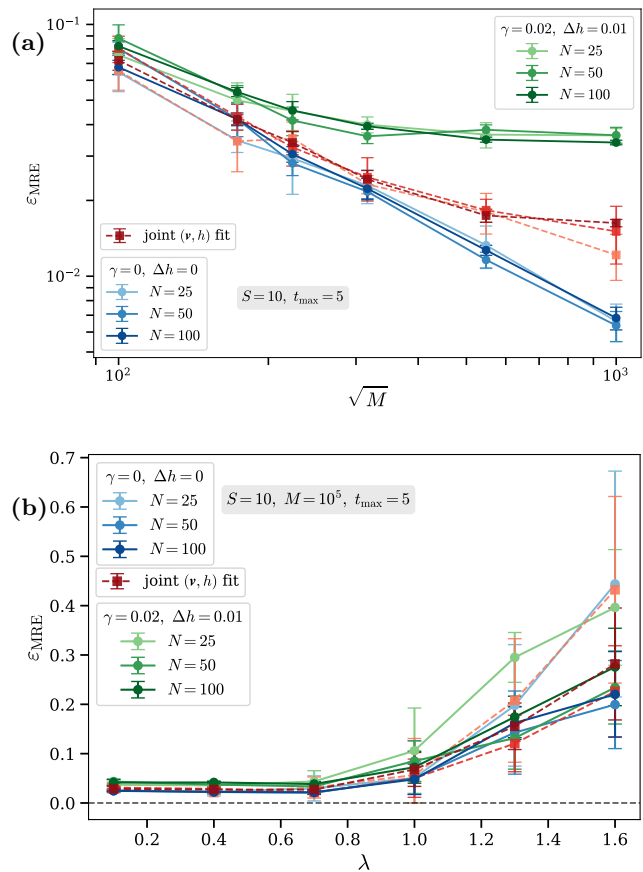

We now turn to arguably the most important question. Can we find the optimal solution in a reasonable amount of time and how does the effort and error scale with system size?

When the experimenter tries to implement a target potential $v_i$, natural limitations of the setup such as diffraction or calibration errors in the digital micromirror device (DMD) inevitably lead to deviations from the intended pattern. The realized potential thus takes the form $v_i + \lambda\,\Omega_i$, where each $\Omega_i$ is drawn independently from a uniform distribution on $[-1,1]$, and $\lambda > 0$ quantifies the calibration error.
%This naturally raises the question of \emph{robustness}: if the optimizer is initialized from such an imperfect potential rather than the true one, does it still converge to the correct global minimum?

To test whether potentials can be scalably inferred from experimental data, we initialize the optimizer from the intended potential $v_i$, which differs from the true potential due to the introduced perturbation. \Cref{fig:MRE_M}(a) shows how the mean reconstruction error $\varepsilon_{\mathrm{MRE}}$ scales with the total number of measurements $M$ and system size $N$ for fixed $S=10$ and $t_{\max}$. The error decreases systematically with increasing $M$, following the expected statistical scaling $\varepsilon_{\mathrm{MRE}}\!\sim\!1/\sqrt{M}$, and remains largely independent of $N$, confirming the scalability of the protocol. In the absence of imperfections in the experimental setup $(\gamma,\Delta h)=(0,0)$, the optimizer converges to the minimum corresponding to the true potential configuration. When state preparation and hopping errors are present, convergence is still achieved but saturates at a small bias. As before, joint optimization over both the on-site potentials and hopping amplitudes further reduces this bias, improving overall reconstruction accuracy. \Cref{fig:MRE_M}(b) presents the dependence of $\varepsilon_{\mathrm{MRE}}$ on the initialization uncertainty $\lambda$. Convergence to the true potential is maintained for $\lambda\!\lesssim\!1$ and does not deteriorate with system size, demonstrating that the protocol is robust. For larger $\lambda$, the error increases as the initial guess departs significantly from the true potentials. This does not mean that the estimation of the potential has failed. The global minimum of the cost function can still be found by trying a large number of initial guesses, but in the worst case, the number of initial guesses required could scale badly with system size.
If this situation arises, there may exist techniques that exploit the finite light cone of the evolution, which could be use to infer the potentials locally.

\section{Implication and Outlook}\label{sec:IO}

Our framework serves as a scalable Hamiltonian potential-learning tool for calibration when only limited experimental data is available. It can be implemented using existing cold-atom technology with global quenches, tunable interactions, and site-resolved imaging. Crucially, we show that the method achieves convergence to the global minimum and remains robust even in the presence of state-preparation errors and uncertainties in the hopping amplitude.  

Interestingly, for fixed number of snapshots $M$ and preparation-error strength $\gamma$, we find that the mean reconstruction error is independent of $N$. For instance, we consistently achieve a mean reconstruction error $\leq5\%$ with a total of about $3\times10^{4}$ snapshots for systems up to 100 sites, which for a cycle time of 10s can be completed in 8 hours. The accuracy and sample efficiency of the method make it suitable for benchmarking many-body localization and randomized protocols, paving the way toward high-fidelity quantum simulation environments. In this work, we only used the particle density inferred from the snapshots as input data for the optimizer. In principle, access to higher-order correlation functions during the time evolution could provide additional information about the state, potentially enabling further reductions in sample complexity~\cite{Valenti_2022}.

Gradient-based optimization routines are often sensitive to the choice of the initial guess. When the loss landscape contains flat regions or multiple local minima, convergence can be hindered, leading to inaccurate reconstructions if the initialization lies far from the true solution.
A potential approach to mitigate these challenges is to introduce a tunable global laser amplitude $\delta$, which controls the overall strength of the applied on-site potential $V$.
Numerical tests and analytical arguments indicate that the global minimum can be found by gradually increasing $\delta$ from near zero. At each step, the optimizer is initialized with the solution obtained at the previous $\delta$. This strategy effectively narrows the feasible solution space and keeps the optimizer within the basin of attraction of the global minimum, at the expense of increasing the sample complexity by the number of steps in the protocol. While potentially useful, we have chosen not to analyze this scheme in detail here, as we expect that in a practical setting, one would prefer to invest more classical computing resources rather than to extend the time scale of the experiment by an order of magnitude.

Extending this approach to interacting systems and time-dependent potentials remains an important direction for future work. One potential application of our protocol is as a Hamiltonian-driven shadow-calibration method~\cite{Huang_2020}, drawing inspiration from Refs.~\cite{Chen_2021,onorati2024noisemitigatedrandomizedmeasurementsselfcalibrating}. It is particularly compatible with generalized shadow-tomography protocols based on Gaussian free fermions and aligns with various Hamiltonian-driven schemes for measuring physical properties through randomized quenches in analog systems~\cite{Naldesi2023,Hu_2022,Tran_2023,mark2024efficientlymeasuringdwavepairing,McGinley_2023,Zhou_2024,Castaneda_2025,Hu_2025,Huang_2020,Denzler_2024,Akhtar_2023,wilkens2024benchmarkingbosonicfermionicdynamics,Impertro_2024,Elben_2018,Shaffer_2021,Chen_2021,Mark_2023,Choi_2023,Poggi_2020,Koh_2022,Brieger_2025,PhysRevResearch.6.043118}. Therefore, this protocol also serves as a robust analog-compatible shadow-tomography scheme that numerically learns the true inversion map for Gaussian free fermions. 
\vspace{1em}
% Self-calibrating classical-shadow protocols have been developed for gate-independent noise~\cite{onorati2024noisemitigatedrandomizedmeasurementsselfcalibrating,Chen_2021,Koh_2022}, where the shadow map factorizes~\cite{Tran_2023} and the calibration map can often be obtained analytically due to the representation-theoretic structure of the Clifford group.

% In contrast, gate-dependent noise poses a greater challenge since no such simplification exists. Our protocol can be interpreted as an analog counterpart that addresses this gate-dependent structure in scenarios where only snapshot data (such as atom occupations) are available.

\section*{Acknowledgements}
We are grateful to Max McGinley and Matt Kiser for collaboration in a related project, and to Daniel Stilck França, Albert H. Werner, Andreas Elben, and Rune Thinggaard Hansen for additional stimulating discussions.
We acknowledge financial support by the Novo Nordisk Foundation under grant numbers NNF22OC0071934 and NNF20OC0059939.

\bibliography{references}

\end{document}